\documentclass[final,3p,times,12pt]{elsarticle}

\usepackage{amssymb}
\usepackage{amsthm}

\usepackage{subfigure}          

\usepackage{color}
\usepackage{lineno}

 \makeatletter
\def\ps@pprintTitle{%
  \let\@oddhead\@empty
  \let\@evenhead\@empty
  \def\@oddfoot{\reset@font\hfil\thepage\hfil}
  \let\@evenfoot\@oddfoot
}

\journal{Astroparticle Physics}

\begin{document}


\begin{frontmatter}


\title{Muon production and string percolation effects in cosmic rays at the highest energies}

 \author[IGFAE]{J. Alvarez-Mu\~niz }
 \author[LIP]{L. Cazon}
 \author[LIP,IGFAE]{R. Concei\c{c}\~{a}o\corref{cor1}}
\ead{ruben@lip.pt}
\cortext[cor1]{Corresponding author}
 \author[CENTRA,IST]{J. Dias de Deus}
 \author[IGFAE]{C. Pajares}
 \author[LIP,IST]{M. Pimenta}
\address[LIP]{LIP, Av. Elias Garcia, 14-1, 1000-149 Lisboa, Portugal}
\address[IST]{Departamento de F\'{i}sica, IST, Av. Rovisco Pais, 1049-001 Lisboa, Portugal}
\address[CENTRA]{CENTRA, IST, Av. Rovisco Pais, 1049-001 Lisboa, Portugal}
\address[IGFAE]{Depto. de F\'{i}sica de Part\'{i}culas \& Instituto Galego de F\'{i}sica de Altas Enerx\'{i}as, Universidade de Santiago de Compostela, 15782 Santiago de Compostela, Spain}

\begin{abstract}
Ultra High Energy Cosmic Rays with energies above $\sim 10^{18}$ eV provide an unique window to study hadronic interactions at energies well above those achieved in the largest man-made accelerators. We argue that at those energies string percolation may occur and play an important role on the description of the induced Extensive Air Showers by enhancing strangeness and baryon production. This leads to a significant increase of the muon content of the cascade in agreement with recent data collected at UHECR experiments. In this work, the effects of string percolation in hadronic interactions are implemented in an EAS code and their impact on several shower observables is evaluated and discussed.
\end{abstract}


\end{frontmatter}


\section{Introduction}
\label{sec:Intro}

Ultra High Energy Cosmic Rays (UHECRs - with energies above $\sim 10^{18}$ eV), are the most energetic particles known in the Universe \cite{Nagano_Watson}. 
These rare subatomic particles reach the Earth and interact with the nuclei in the atmosphere initiating huge cascades of lower energy particles usually 
referred in the literature as Extensive Air Showers (EAS). The origin and composition of the UHECRs are still unknown and are an intense research 
subject \cite{Olinto_UHECR}. Their interactions in the atmosphere are also very interesting, as the centre-of-mass energy of the collisions reaches 
some hundreds of TeV, well above the energies reached at the largest man-made accelerators to study particle physics. 
This unique physics window can be indirectly explored in large arrays devoted to UHECR detection, of which the 
Pierre Auger Observatory at Malarg\"{u}e, Argentina \cite{PAO} is the paradigmatic example.

Recent results on the muon content of EAS at ground level obtained at the Pierre Auger Observatory, have demonstrated that 
the simulations of EAS show a significant deficit in the number of muons predicted when compared to data \cite{muonAuger,muonAuger_inclined}.
The relative deficit is $\sim$50\% when data are compared to simulations of proton-induced showers with the QGSJ{\textsc{et}}-II 
model. This discrepancy cannot be explained by the unknown UHECR composition alone. 
On the other hand, with data collected at the KASCADE-Grande experiment \cite{muonKASCADE} it has been shown that the muon content at ground level 
up to energies of $\sim10^{17.7}$ eV is, as expected, between the number of muons predicted in proton and iron initiated showers, regardless of the 
high energy interaction model being used. 
This scenario would necessarily imply a transition in the behaviour of the hadronic interactions in the energy region between the highest energy 
events in the KASCADE-Grande array ($\sim10^{17.7}$ eV) and the lowest energy of the events used for muon analysis at the Pierre Auger 
Observatory ($\sim10^{18.4}$ eV).

The increase of the number of muons in simulations of EAS can be achieved by effectively reducing the amount of shower energy that is transferred to the electromagnetic component, which is mainly attained by the decay of neutral pions \cite{Matthews}.

The string percolation model (SPM) has been quite successful in describing global features of data in AA collisions at RHIC (with energy up to $200$ GeV) 
and in pp collisions at LHC (with energy up to $7$ TeV for the moment)
\cite{PercNucData1,PercNucData2,PercNucData3}. In the model QCD strings are formed, stretching between beam and target. The strings may interact by fusion 
of clusters of overlapping strings in the transverse plane of the collision, in a situation similar to two-dimensional percolation 
theory \cite{PurePerc0,PurePerc1,PurePerc2}. The predictions of the SPM are: i) a reduction of the height of the rapidity plateau around $y=0$; 
ii) an increase in the transverse momentum $p_T$ of the particles produced in the collisions due to string interactions; 
iii) heavy quark production as a consequence of clustering and percolation. 
In this paper we argue that string percolation effects may be found at very high energies ($\sqrt{s} \sim 40$ TeV) in UHECR data.
 Percolation effectively results in an enhancement of kaon and baryon production which results in the reduction of the amount of 
energy being transferred to the electromagnetic cascade in the high energy interactions.
It is important to note that 
the percolation model does not forbid some other phenomena such as a rapid increase of the primary-air cross section, 
as for instance described in \cite{BlackDisc}. This would actually drive the behaviour of the momenta of the 
distribution of the depth of maximum of the electromagnetic component ($X^{\rm e.m.}_{\rm max}$) of the shower. 

The paper is organized in the following way: in Section \ref{sec:theory} the string percolation model is presented and its impact 
on the hadronic interaction physics is discussed; afterwards (in Section \ref{sec:Implementation}) we explain how the percolation 
effects were implemented in an EAS code; in Section \ref{sec:results} the effect of percolation on shower observables, in particular 
the depth of shower maximum and the number of muons at ground, is evaluated and discussed. Some final remarks and conclusions are given 
in Section \ref{sec:Conclusions}.

\section{String Percolation Model}
\label{sec:theory}

In the string model (see for instance \cite{DPM,PRL15}), colour strings stretched between the beam and the target are formed in high energy collisions. 
In the impact parameter (transverse) plane of a collision the strings appear as disks with a characteristic transverse radius $r_0$. 
The disks may overlap, cluster and fuse (percolate). The relevant parameter in the String Percolation Model (SPM) is the so-called 
transverse string density $\eta$ defined as \cite{PRL12},

\begin{equation}
\eta \equiv \left( \frac{r_0}{\overline{R}} \right)^2 \overline{N}_s
\label{eq:eta}
\end{equation}
where $\overline{R}$ is the effective radius of the interaction area in the transverse plane, and $\overline{N}_s$ is the average
 number of strings formed in the collision, which typically grows with the centrality and energy of the collision \cite{Ns_PRL}. 
Depending on the value of $\eta$ the strings may occupy a significant fraction of the transverse area of the collision, overlap, 
cluster and percolate. If $\eta \le 1$ the average number of strings per cluster $\left< N \right>$ is approximately one 
(the strings do not overlap), while if $\eta \gg 1$ then $\left< N \right> \approx \overline{N_s}$ (and the majority of the strings overlap). 
It is then important to relate the average number of strings per cluster $\left< N \right>$ to the average area $\left< A \right>$ 
occupied by a cluster. Following \cite{PRL14} $\left<A\right>$ is written as,
\begin{equation}
\left< N \right> =  \left< A \right> \frac{\eta}{1 - e^{-\eta}}
\label{eq:N}
\end{equation}
with $\left< A \right>$ given in units of $r_0$ by,
\begin{equation}
\left< A \right> =  f(\eta) \left[  \left( \frac{\overline{R}}{r_0} \right)^2 (1-e^{-\eta}) -1 \right] +1,
\label{eq:A}
\end{equation}
where $f(\eta)$ is a percolation function \cite{Ns_PRL,PRL14}, 
\begin{equation}
f(\eta) = (1+e^{-(\eta-\eta_c)/a})^{-1}.
\label{eq:feta}
\end{equation}
$f(\eta)$ characterizes the transition from individual to percolated strings, controlled by the parameters $\eta_c$, 
the critical transverse string density, and $a$, the parameter controlling the slope of the curve at the transition point. 
Comparison with Monte Carlo simulations \cite{PRL14} gives $\eta_c = 1.15$ and $a = 0.07$ in the case 
of a uniform proton density profile as a function of the impact paramater. 
For a Gaussian profile the parameters would be $\eta_c = 1.59$ and $a = 0.14$. 

One should note that, contrary to $\overline{N_s}$, the average number of strings per cluster, 
and the quantity $\left< A \right>$ are sensitive to the percolation transition.
If $\eta\rightarrow 0$ then  $\left< A \right>\rightarrow 1$ (in units of $r_0$) and
when $\eta\rightarrow\infty$ then  $\left< A \right>\rightarrow (\overline{R}/r_0)^2$.

In order to relate the average number of formed strings $\overline{N_s}$ to the collision energy $\sqrt{s}$, we use the 
parameterisation in \cite{Ns_PRL} and write,
\begin{equation}
\overline{N}_s = b + (2-b) \left( \frac{s}{s_0} \right)^{\lambda}
\label{eq:Ns_PRL}
\end{equation}
where the free parameters $b = 1.37$, $\sqrt{s_0} = 10$ GeV and $\lambda = 0.23$ have been 
adjusted to experimental data as explained in \cite{Ns_PRL,Xsec}.

\begin{figure}[htbp]
\begin{center}
\includegraphics[width=0.75\textwidth]{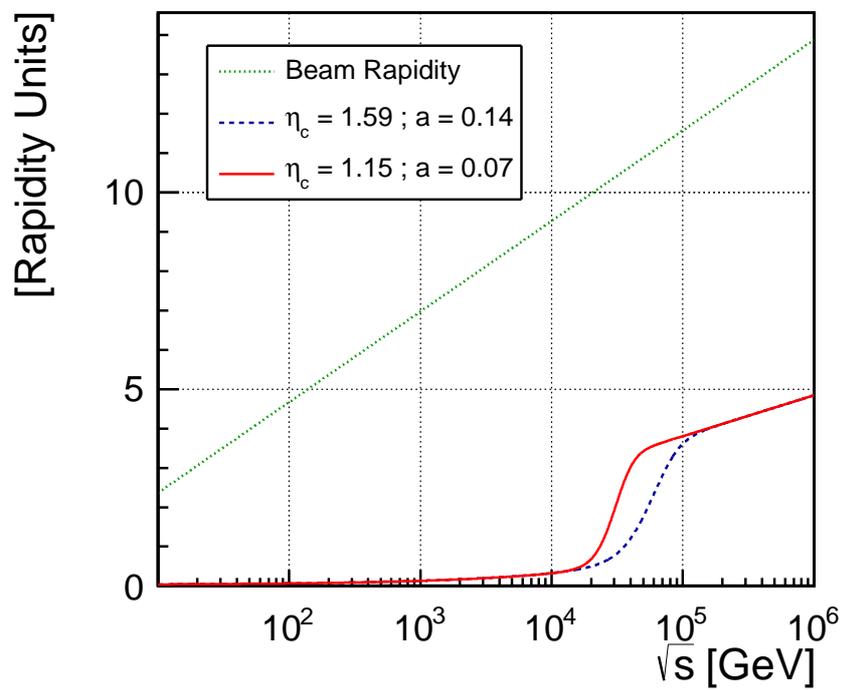}
\caption{Length of the rapidity plateau in a high energy collision $\Delta Y$ (in rapidity units) as a function of $\sqrt{s}$. 
The red line is the result for a proton with a uniform density profile distribution, while the blue line corresponds to a Gaussian profile. 
The beam rapidity is also shown in green for reference. All the results are displayed only for the positive (forward) rapidity region
of the collision.}
\label{fig:stringYincrease}
\end{center}
\end{figure}

According to the string percolation model, when strings in a collision fuse the length of the \emph{rapidity plateau} $\Delta y$ increases in proportion to the number of clustering strings. Following the treatment in \cite{Ns_PRL},
\begin{equation}
\Delta y \sim 2 \log \left( \left< N \right> \right)
\label{eq:stringYincrease}
\end{equation}
In Fig.~\ref{fig:stringYincrease}, the increase in the length of the rapidity plateau as a function 
of $\sqrt{s}$ given in Eq.~(\ref{eq:stringYincrease}) 
is plotted for the forward region of the collision only. The beam rapidity is also plotted for comparison.

Since the strings are stretched in rapidity, energy conservation implies a natural reduction of the hadronic interaction multiplicity. 
The most forward and backward regions of the collision remain untouched, i.e., the rapidity of the strings there is not increased.
The reduction of multiplicity is realized through a colour summation factor $F(\eta)$ decreasing with $\eta$ and given by \cite{PRL15,PRL16},
\begin{equation}
F(\eta) = \sqrt{\frac{1-e^{-\eta}}{\eta}}
\label{eq:Feta}
\end{equation}
%

In the spirit of the Schwinger mechanism for particle production \cite{Schwinger} the probability of production of a $q\bar{q}$ pair (meson) in
a string is,
\begin{equation}
P_{q\bar{q}} \propto e^{\frac{- \pi m_q^2}{k_0}}
\label{eq:schwinger}
\end{equation}
where $m_q$ is the corresponding quark mass and $k_0 \simeq 0.2$ GeV$^2$ is a phenomenological parameter related to the effective color field, 
the so-called \emph{string tension} \cite{ktension}. If we compare the probability of producing a heavy quark pair $P_{Q\bar Q}$
(such as a strange quark pair for instance),
to the production of a light quark pair $P_{q\bar q}$ (such as a $u\bar u$ pair for instance), we obtain, in the absence of percolation: 
\begin{equation}
P_{su}(\eta=0)=\frac{P_{s\bar{s}}}{P_{u\bar{u}}}(\eta=0) \sim \exp \left[-\frac{\pi}{k_0} \left( m_s^2 - m_u^2  \right) \right]
\label{eq:schwinger2}
\end{equation}

When string percolation is effective the string tension increases due to the presence of higher color charges associated to the clusters.
It is natural to associate the increase of the string tension to the increase of the number of strings per cluster corrected by the color 
reduction factor $F(\eta)$ in Eq.~(\ref{eq:Feta}). With percolation we have an increased string tension $k$ given by,
\begin{equation}
k = k_0 \left< N \right> F(\eta)
\label{eq:kstring}
\end{equation}
As a consequence the probability of producing a $s\bar s$ pair relative to the probability of producing a $u\bar u$ pair is given by, 
%
\begin{equation}
\frac{P_{s\bar{s}}}{P_{u\bar{u}}}(\eta) \sim \exp \left[-\frac{\pi}{k} \left( m_s^2 - m_u^2  \right) \right]
\label{eq:schwinger2}
\end{equation}
which can be cast as,
\begin{equation}
\frac{P_{s\bar{s}}}{P_{u\bar{u}}}(\eta) \sim \exp \left[-\frac{\pi}{k_0 \left< N \right> F(\eta)} \left( m_s^2 - m_u^2  \right) \right] = 
P_{su}(0)^{\left[ \left<N\right>F(\eta) \right]^{-1}}
\label{eq:strangeness}
\end{equation}
where $P_{su}(0)$ is the probability of producing a quark s relative to a quark u without percolation. 
In the hadronic interaction models 
this quantity is of the order of $0.1$, in particular in the QGSJ{\textsc{et}}-II model $P_{su}(0) = 0.07$. 

\begin{figure}[htbp]
\begin{center}
\includegraphics[width=0.75\textwidth]{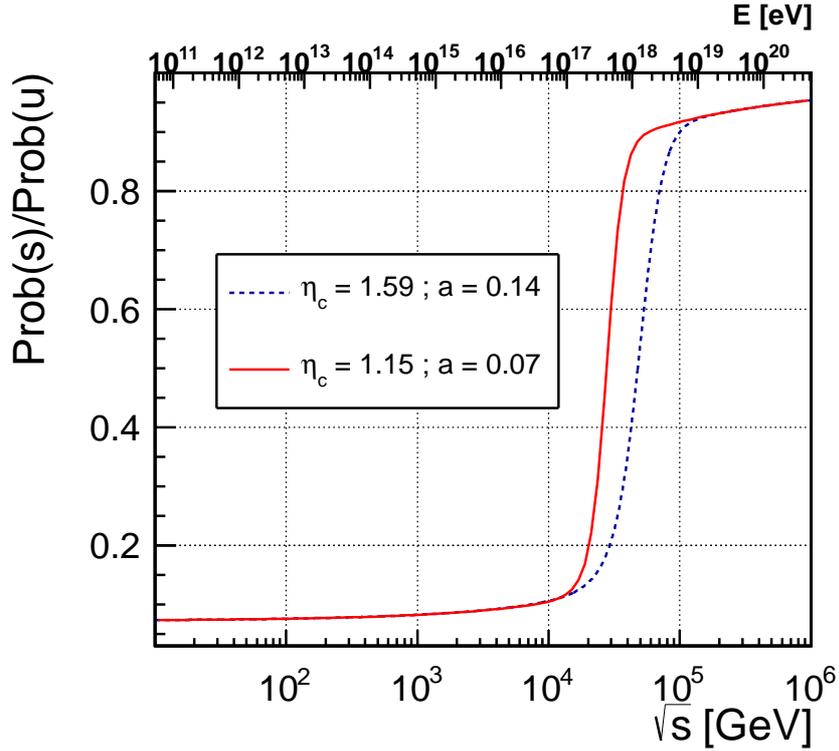}
\caption{Probability of producing a $s\bar{s}$ relatively to $u\bar{u}$ pair as a function of $\sqrt{s}$
in the QGSJ{\textsc{et}}-II model accounting for string percolation effects. The red line is for a proton 
with an uniform density profile while the blue line is assuming a Gaussian profile. The upper scale 
is the equivalent energy of a proton in the laboratory frame of the Earth.}
\label{fig:pquarkperc}
\end{center}
\end{figure}

Combining Eqs.~(\ref{eq:eta}-\ref{eq:Ns_PRL}), (\ref{eq:Feta}) and (\ref{eq:strangeness}) the evolution with energy 
of the probability of producing strangeness 
in QGSJ{\textsc{et}}-II as a function of $\sqrt{s}$ can be obtained. The ratio $P_{s\bar{s}}/P_{u\bar{u}}$ is shown in Fig.~\ref{fig:pquarkperc}. 
The string percolation model predicts a transition in the center-of-mass energy region of $\sqrt{s}\sim 20-30$ TeV, above LHC energies, 
and typically below the energies of the events collected at the Pierre Auger Observatory 
(the energy of Auger events used for muon analysis is above $E = 10^{18.4}$ eV, which in the center of mass of proton-proton collisions corresponds to $\sqrt{s} \simeq 68$ TeV).

Similarly to the production of $q\bar q$ pairs, in the string percolation model the probability of producing a 
baryon (associated to the diquark production, $qq\overline{qq}$) gets increased in essentially the same way, due to the increase of the string tension of the fused string. As a consequence an increase in the number of baryons is expected. 



The parameters of the model are presented in Table~\ref{tab:parameters}. The two first parameters, $\lambda$ and $r/\overline{R}$, are given by the experimental accelerator data and the best fit values are displayed in the column designated \emph{Preferred Value}. The parameters allowed region, due to the experimental uncertainties are also shown. The remaining two parameters, $\eta_c$ and $a$, are obtained through simulations and the presented values reflect the uncertainty on the proton density profile as a function of the impact parameter. Two extreme case are taken: Gaussian and uniform distribution.
\begin{center}
\begin{table}[htbp]
\begin{center}
\begin{tabular}{c|ccc}
\hline
Parameter & Min Value & Max Value & Preferred Value\\
\hline
$\lambda$ & 0.21 & 0.28 & 0.23 \\
$r/\overline{R}$ & 0.2  & 0.3  & 0.2 \\
$\eta_c$ & 1.15 & 1.59 & $\--$ \\
$a$ & 0.07 & 0.14 & $\--$ \\
\hline
\end{tabular}
\caption{Main parameters of the string percolation model (see text for definitions).}
\label{tab:parameters}
\end{center}
\end{table}
\end{center}

\section{Percolation of strings in Extensive Air Showers}
\label{sec:Implementation}

To study the impact of percolation of strings in high energy collisions on the properties of UHECR-induced showers,
 we used the EAS simulation code CONEX \cite{Conex1,Conex2}. CONEX is a hybrid Monte Carlo for the study of EAS. 
In CONEX the most energetic interactions, those more relevant for the shower development, are simulated explicitely 
with the Monte Carlo, while the remaining interactions are treated by solving cascade equations \cite{Conex1,Conex2}. 
This makes it a very fast code allowing the simulation of large statistical samples of EAS. 
However, with CONEX only information on the shower longitudinal development along shower axis ($X_{\rm max}$, 
number of particles at a given depth) can be obtained, 
and there is no information on the lateral spread of the shower particles in the direction perpendicular to the shower axis.  
Despite this, CONEX is the ideal tool for studying the effect of percolation on the average behaviour 
of some of the main shower observables, and their dependence on shower energy - a task that naturally increases 
the amount of EAS simulations needed and calls for a fast and flexible simulation of EAS. 

The hadronic interaction model used as a baseline for the study of the influence of percolation of strings on EAS development 
is the QGSJ{\textsc{et}}-II.03 model \cite{QGSII1,QGSII2}. 
The percolation mechanism was implemented in QGSJ{\textsc{et}}-II.03 following two steps: 
the probability of heavy quark production was enhanced according to Eq.~(\ref{eq:strangeness}); 
and the rapidity of the strings was increased following Eq.~(\ref{eq:stringYincrease}). 
The latter step leads to a reduction of multiplicity in the collision as will be explained below.
All these changes were done in a way that does not compromise the consistency of the hadronic interaction model. 
This means that the modifications to QGSJ{\textsc{et}}-II.03 were done either on input parameters of the code (the increase of 
the probability of heavy quark production), or by applying {\it ad-hoc} changes to the particles produced by the model after a hadronic interaction 
(as done to emulate the behaviour of the increase of string rapidity and the reduction of multiplicity). 
All the hadronic interaction features described by the model remain untouched.
Below we give more details on how these changes were implemented.

\begin{figure}[htbp]
\begin{center}
\includegraphics[width=0.45\textwidth]{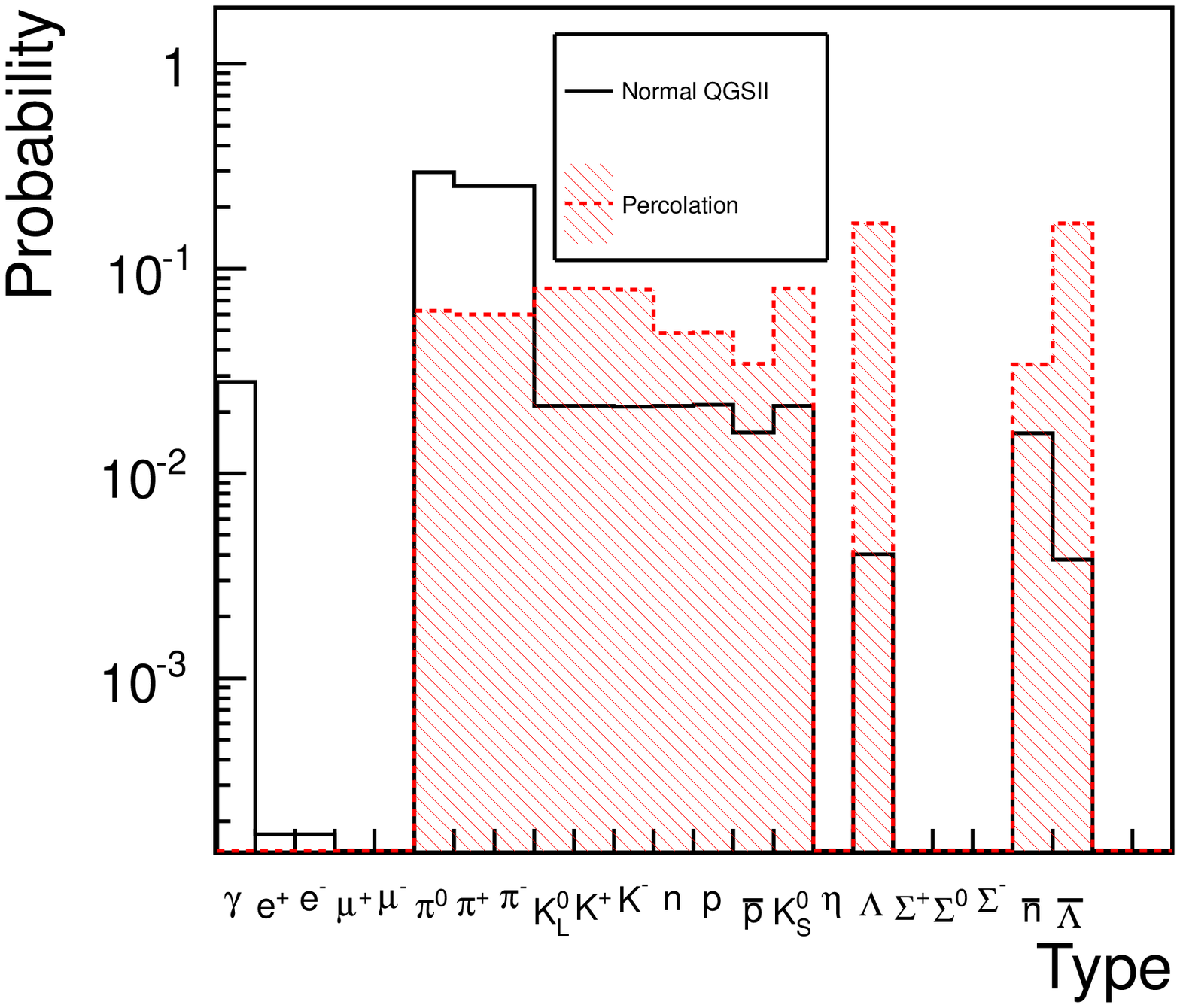}
\includegraphics[width=0.45\textwidth]{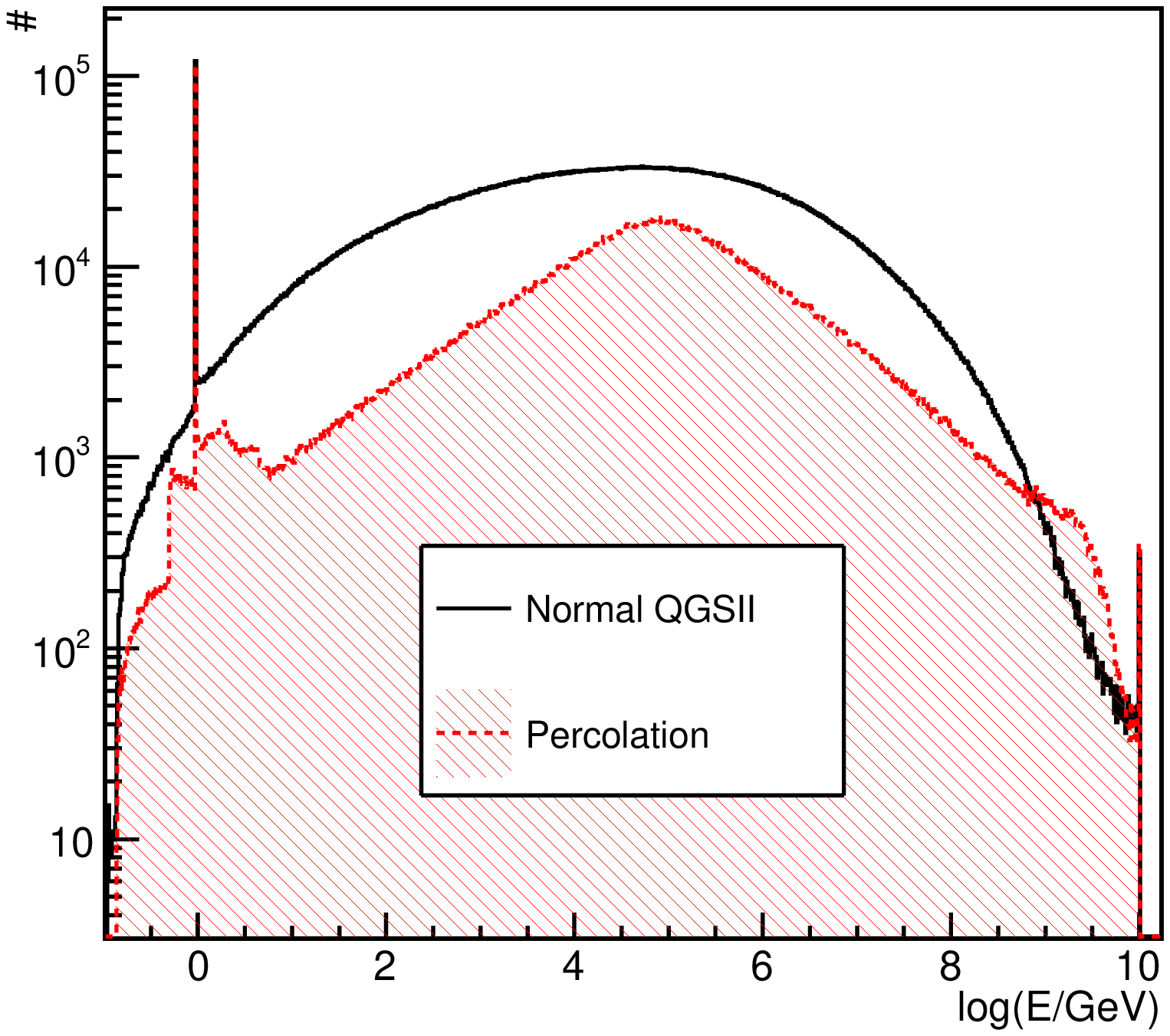}
\caption{Left panel: Probability of particle production as a function of the type of particle as 
predicted with the QGSJ{\textsc{et}}-II.03 model without percolation (``Normal QGSII") and after 
enhancing the probability of strange quark production as expected in the percolation model and explained in the text. 
Right panel: Energy spectrum of the produced particles as predicted with the 
QGSJ{\textsc{et}}-II.03 hadronic model without percolation and after increasing the rapidity of the strings
as expected in the percolation model and explained in the text.  
In both cases proton-air collisions at $E = 10^{19}$ eV were simulated.}
\label{fig:conexEspectrum}
\end{center}
\end{figure}

The probability of producing strange quark pairs with respect to light quarks pair is one of the input parameters in the QGSJ{\textsc{et}}-II.03 model. 
In the original code, this parameter does not depend on the energy of the collision and is fixed to $P_{su}(0)=0.07$. 
To implement the effects of percolation of strings, the value of $P_{su}$ was modified using Eq.~(\ref{eq:strangeness}).
As explained before, $P_{su}$ depends on the interaction energy and this dependence is plotted in Fig.~\ref{fig:pquarkperc}.
As argued in Section~\ref{sec:theory}, the increase of the string tension should also affect the probability of diquark production 
(and hence the production of baryons), and in fact we have assumed the same evolution with $\sqrt{s}$ for the enhancement of the probability 
of diquark pair creation. The impact of such modifications on particle production in high energy hadronic interactions (at a reference energy of 
$\sqrt{s} \sim 100$ GeV) can be seen in the left panel of Fig.~\ref{fig:conexEspectrum}. As expected there is a notorious increase in the 
number of strange mesons being produced, in particular kaons, as well as in the production of baryons with respect to the QGSJ{\textsc{et}}-II.03 model without
string percolation. Electromagnetic particles and particles responsible for the production of electromagnetic particles (mainly neutral pions) are suppressed. 
This is relatively important for the electromagnetic component of the UHECR-induced showers since the decay of $\pi^0$s is the main channel that feeds 
this shower component.  

The output of each hadronic interaction is obtained with the QGSJ{\textsc{et}}-II.03 model without percolation. 
The resulting particles are then processed in order to emulate the effects of percolation on 
the increase of string rapidity. For this purpose we have worked in the collision centre-of-mass reference frame and 
particles are then boosted to the laboratory frame and injected into the CONEX Monte Carlo code.
Only the properties of the particles with rapidity $|y| < y_{beam} - \Delta y$, (where $y_{beam}$ is the beam rapidity and $\Delta y$ is the rapidity 
increment in Eq.~(\ref{eq:stringYincrease})), 
were modified to implement the effects of percolation. 
This last condition is necessary to fulfill energy conservation, and at the same time it leaves the leading particle 
(and diffractive particles) untouched, where percolation effects are not expected to play a role. 
Focusing the discussion on positive rapidities 
the average value of the rapidity is increased according to Eq.~(\ref{eq:stringYincrease}). 
The rapidity distribution is divided in $N$ bins and each bin is stretched in rapidity by an amount $\Delta y/N$. 
Afterwards, the particles in each old bin in rapidity are randomly distributed in the new bin so that their rapidity increases on average.
In each bin, local energy conservation is applied: particles in the old bin in $y$ are redistributed into the new stretched 
bin until the total energy of the particles in the old bin is completely spent.
This leads to a natural reduction of the number of particles in each bin, and hence to a reduction 
of the total multiplicity of the collision. 
The final result is a smooth energy distribution that, in the center-of-mass framework, has a higher average energy but less 
particles due to energy conservation. The impact of this implementation on hadronic interactions can be seen in the energy spectrum
of the particles in the laboratory frame shown in the right panel
of Fig.~\ref{fig:conexEspectrum}. Several structures emerge: the peak at lower energies is due to the collision spectators 
(note that these are proton-air collisions), while the peak at the highest energies is attributed to diffractive events. 
Percolation is not expected to change these two features, and they are equal before and after the effects of percolation
were implemented. As the laboratory framework is highly boosted in one direction relative to the centre of mass frame, 
the particles are either migrated to higher or lower energies, depending on their direction with respect to the boost. 
One can also easily see the reduction of multiplicity due to percolation. 
At $E = 10^{19}$ eV this reduction is $\sim 65\%$, i.e. the multiplicity of a hadronic interaction 
with percolation would be in average $\sim 35\%$ of that in a standard interaction.

\section{Percolation effects on shower observables}
\label{sec:results}

In this section we present and discuss the impact on shower observables of the presence of string percolation 
in hadronic interactions at ultra high energies.

\begin{figure}[htbp]
\begin{center}
\includegraphics[width=0.75\textwidth]{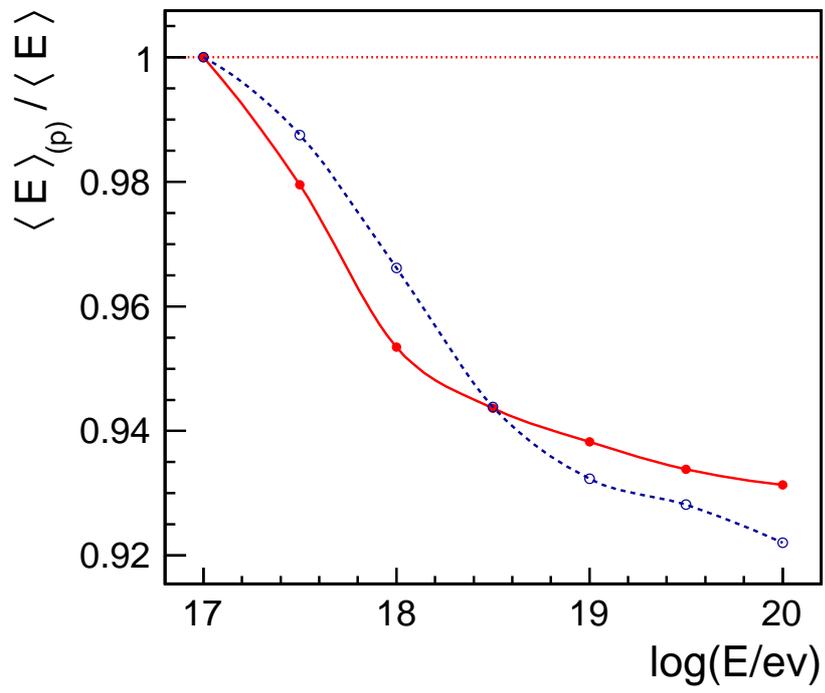}
\caption{Ratio of the average energy transferred to the electromagnetic component of showers with percolation ($\left< E\right>_{(p)}$) 
and without percolation effects (normal QGSJ{\textsc{et}}-II) as function of the energy of the primary proton inducing the shower. 
The red solid line is for a proton with a uniform density profile distribution, while the blue dashed line assumes a Gaussian 
density profile.}
\label{fig:em_energy}
\end{center}
\end{figure}

As pointed out in Section~\ref{sec:Implementation}, one of the main consequences of string percolation in hadronic 
interactions is a reduction of both electromagnetic particles and of particles such as neutral pions that when decaying 
feed the electromagnetic component of the shower (see left panel of Fig.~\ref{fig:conexEspectrum}). The energy in these particles will 
be transferred to the hadronic component of the cascade, and a small fraction of it will contribute to the so called \emph{invisible energy} \cite{invE1,invE2},
defined as the energy in the shower carried by muons and neutrinos. This effect has consequences on the energy calibration 
performed in modern UHECRs experiments, such as the Pierre Auger Observatory \cite{AugerEnergy} 
and the Telescope Array experiment \cite{TAFD}. In both experiments, showers observed simultaneously with 
the fluorescence and the surface detector (typically $\sim 10-15\%$ of the events) serve to calibrate 
the energy of the showers detected with the surface array alone.
In fluorescence detectors, the fluorescence light emitted by the excitation of the nitrogen molecules due to the development 
of the shower through the atmosphere is collected, providing in this way a \emph{quasi}-calorimetric measurement of the primary energy. 
The excitation of the nitrogen molecules is mostly induced by low energy electrons. String percolation should affect 
the energy assignment by the fluorescence detector since, as stated above, the energy transferred to the electromagnetic
particles in the shower decreases when percolation of strings is effective. This effect is shown in Fig.~\ref{fig:em_energy} 
where the ratio of the average electromagnetic energy with percolation, ($\left< E\right>_{(p)}$), and without percolation as obtained in 
Monte Carlo simulations of EAS performed with CONEX, is plotted as a function of the primary cosmic ray energy. 
The ratio of the electromagnetic energy in showers with percolation to standard proton showers, simulated with QGSJ{\textsc{et}}-II, 
is seen to decrease as the primary energy increases. At $E=10^{19}$ eV the energy estimated through the longitudinal development 
of the light collected by the fluorescence detector would be $\sim 5\%$ smaller when percolation is effective, 
inducing in this way a systematic underestimate of the energy scale used to calibrate the events observed with the surface detector alone. 
At lower energies, $E = 10^{17}$ eV, the percolation 
transition has not occurred yet (see Fig.~\ref{fig:pquarkperc}) and the ratio in Fig.~\ref{fig:em_energy} is one by definition. 
In Fig.~\ref{fig:em_energy} the results for two extreme cases of the proton density profile: uniform distribution and 
Gaussian distribution are shown. In both scenarios the trend is the same. 

\begin{figure}[htbp]
\begin{center}
\includegraphics[width=0.45\textwidth]{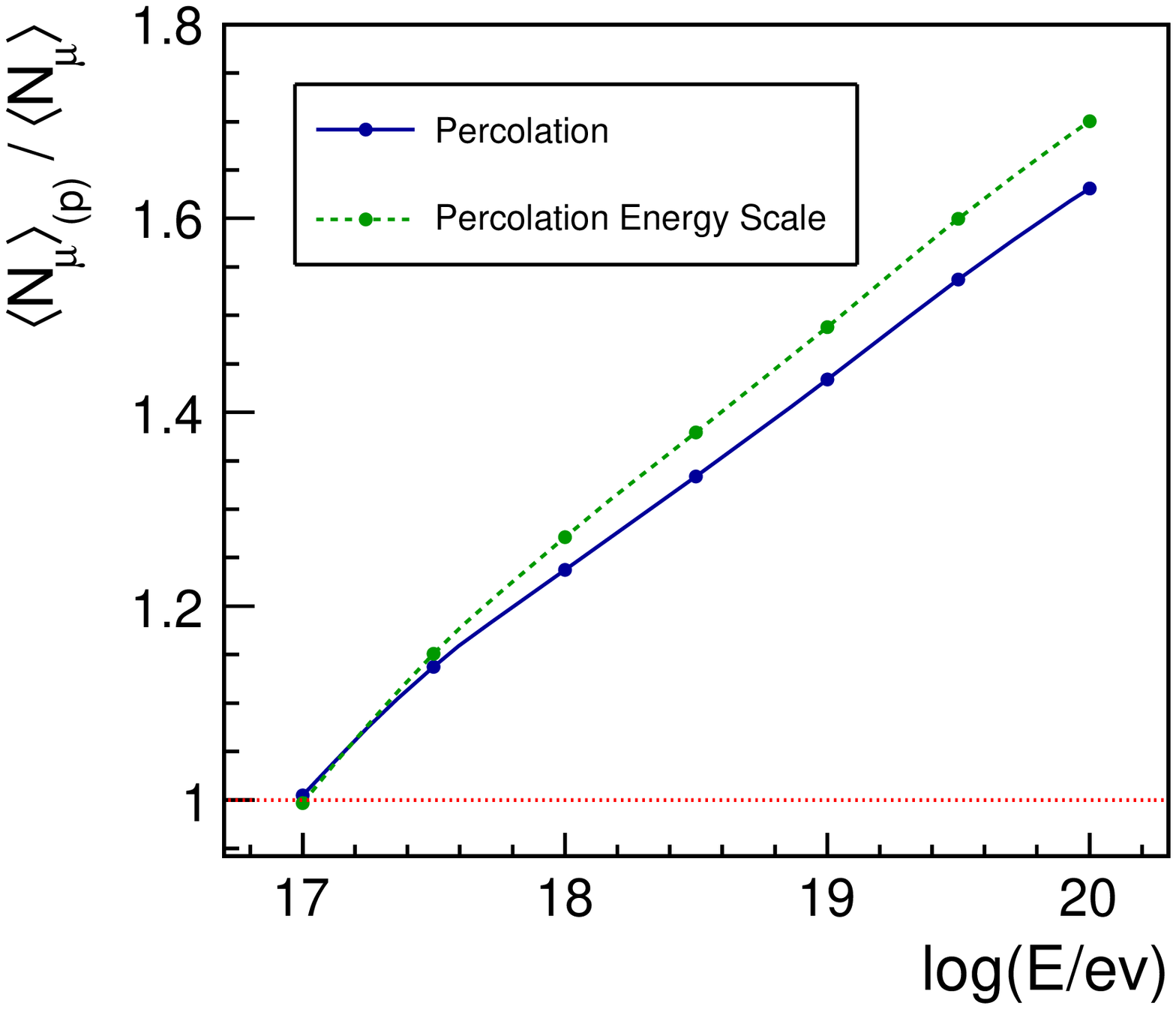}
\includegraphics[width=0.45\textwidth]{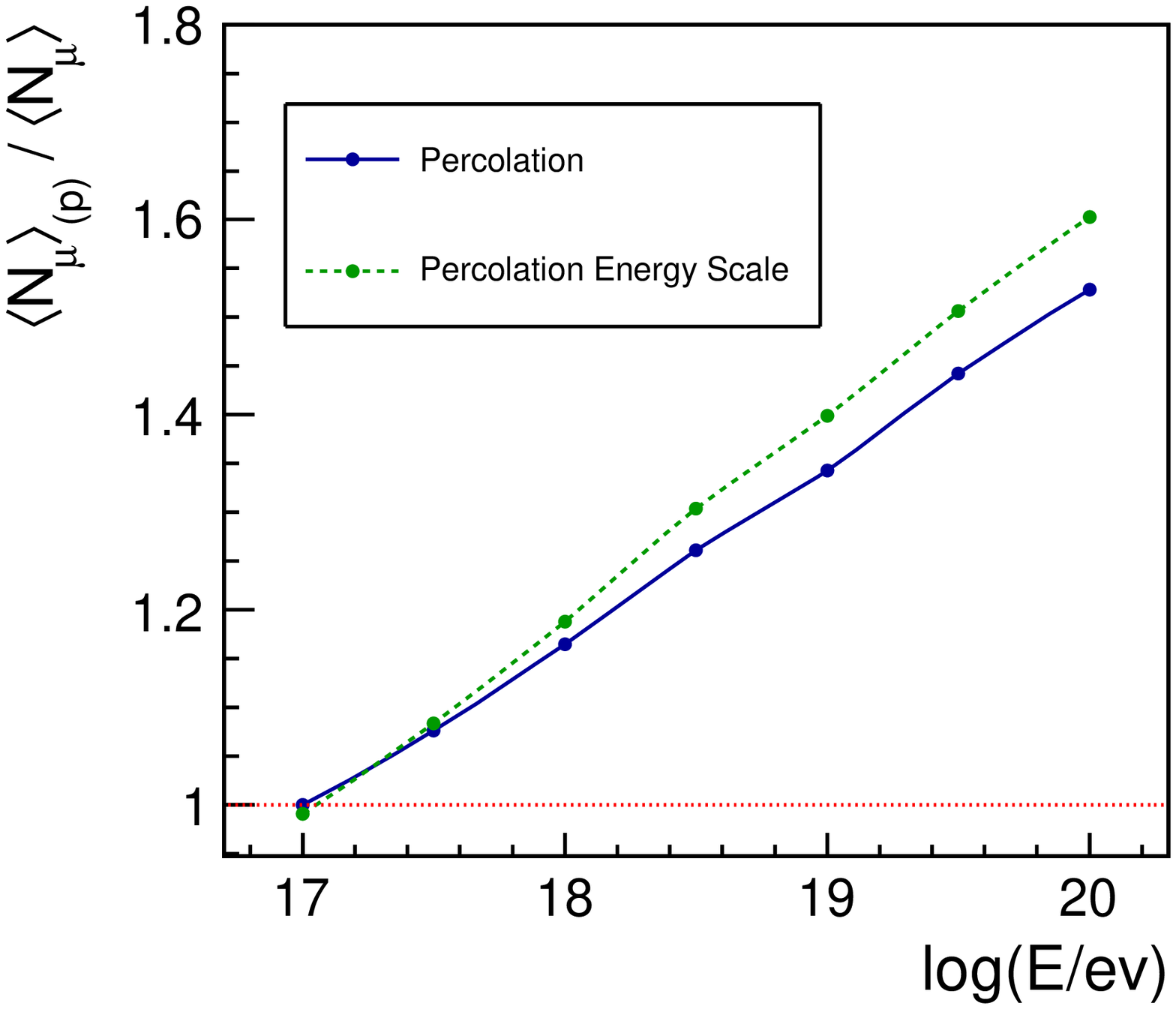}
\caption{Ratio of the average number of muons in showers with percolation ($\left< N_{\mu}\right>_{(p)}$)
and without percolation effects (simulated with the standard QGSJ{\textsc{et}}-II) as function of the energy of the primary proton inducing the shower. 
The left panel corresponds to a uniform density profile distribution, while in the right panel a Gaussian 
density profile is assumed. The impact of the energy scale on the ratio is shown with the green dotted line.}
\label{fig:ratioNmu}
\end{center}
\end{figure}

The impact of percolation on the number of muons at ground level is shown in Fig.~\ref{fig:ratioNmu} where the ratio of the number 
of muons in showers with percolation effects and in standard proton showers is plotted as a function of the primary particle energy. 
As expected, the enhancement on the number of strange mesons and baryons, produced in the highest energy collisions in the shower, 
leads to an increase of the average number of muons at ground level. This effect accounts 
for $\sim 30\%$ of the increase, while the hardening of the particle energy spectrum due to percolation (right panel of Fig.~\ref{fig:conexEspectrum}),
which leads to a more penetrating shower, accounts for an additional increase of $\sim 10\%$.
The total increase on the number of muons at ground level at $E = 10^{19}$ eV is $\sim 34\--43\%$ 
depending on the percolation model assumptions.
The effect increases with the primary energy, both for a uniform density profile of the proton (left panel of Fig.~\ref{fig:ratioNmu}) 
and for a Gaussian proton density profile (right panel). Therefore, the string percolation mechanism, provides a possible explanation
for the deficit of muons in standard simulations observed at ultra-high energies $E > 10^{18}$ eV when compared to experimental data \cite{muonAuger}. 
Moreover, given the energy at which the onset of percolation phenomena is expected, 
it becomes clear why standard hadronic interactions models are able to bracket the lower energy data
observed for instance in the KASCADE-Grande experiment, where a deficit is not seen because string percolation is not effective. 

The reduction of the energy transferred to the electromagnetic component would effectively lead to a lowering of the energy scale, as explained above. 
This means that a number of muons measured at ground level would be attributed to a slightly lower cosmic ray energy than in reality 
if percolation effects are not accounted for when estimating the energy of the event. 
This effect is shown in Fig.~\ref{fig:ratioNmu} (dashed lines) where a shift of the energy scale following 
the trend in Fig.~\ref{fig:em_energy} is applied. As a consequence at $E = 10^{19}$ eV 
the number of muons with respect to standard proton QGSJ{\textsc{et}}-II showers is $\sim 40\--49\%$ higher. 

It is important to notice that the increase in Fig.~\ref{fig:ratioNmu} affects the total number of muons 
arriving at ground level, integrated over all distances to the shower axis. Thus, the features observed 
in Fig.~\ref{fig:ratioNmu} are not expected to be directly comparable to those observed by 
the Pierre Auger Observatory where the number of muons at a distance of 1000 m from the core was measured. 
A direct comparison would require a precise knowledge of the shape of the lateral distribution which is not
possible to obtain in simulations with the CONEX Monte Carlo, 
and would require a full 3D simulation (such as for instance CORSIKA) which is currently out of the scope of this paper. 
Finally, it is also important to remark that we do not intend to argue about the absolute number 
of muons at ground, but only to discuss general trends with energy, as the number of muons still has 
some dependence on several low energy hadronic interaction phenomena which are not yet fully understood \cite{NA61}.

\begin{figure}[htbp]
\begin{center}
\includegraphics[width=0.75\textwidth]{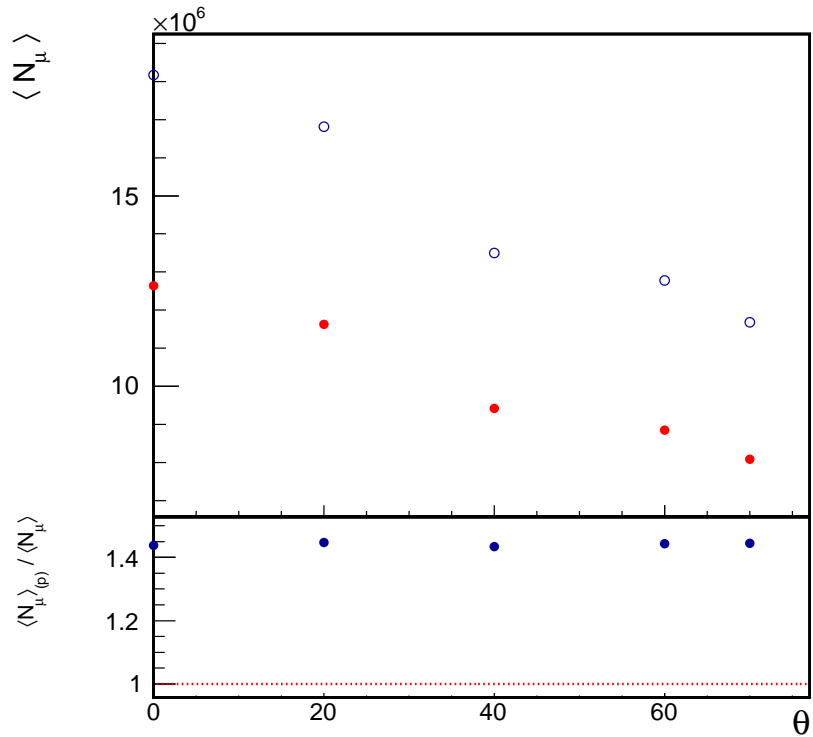}
\caption{Upper panel: evolution of the number of muons with the shower zenith angle $\theta$. Blue open dots: showers with percolation.
Red filled dots: showers without string percolation. In both cases proton induced showers at $E = 10^{19}$ eV were simulated. 
Bottom panel: ratio of the average number of muons at ground level with and without percolation. 
These results were obtained assuming a uniform density profile for the proton.}
\label{fig:muon_vs_zenith}
\end{center}
\end{figure}

The dependence of the total number of muons at ground level with energies above $1$ GeV, with the shower zenith angle $\theta$, 
was also investigated and the results are displayed in Fig.~\ref{fig:muon_vs_zenith}. In the upper panel 
of Fig.~\ref{fig:muon_vs_zenith} the number of muons as a function of $\theta$ is displayed for standard
showers (without string percolation) and for showers with string percolation effects included.  
The number of muons is larger in showers with string percolation effects as explained before and seen in Fig.~\ref{fig:ratioNmu}, 
but in both cases (percolation and standard) the total number of muons at ground level follows the same trend and decreases 
with $\theta$. 
In fact, the ratio between the number of muons in showers with percolation and standard showers is constant
with $\theta$ as shown in the lower panel of Fig.~\ref{fig:muon_vs_zenith}. 
These plots were done for the case of an uniform density profile of the proton but the same conclusion can be withdrawn for 
a Gaussian profile. This means that percolation phenomena alters the total 
number of muons at ground, but its evolution with the shower zenith angle remains the same. 

\begin{figure}[htbp]
\begin{center}
\includegraphics[width=0.45\textwidth]{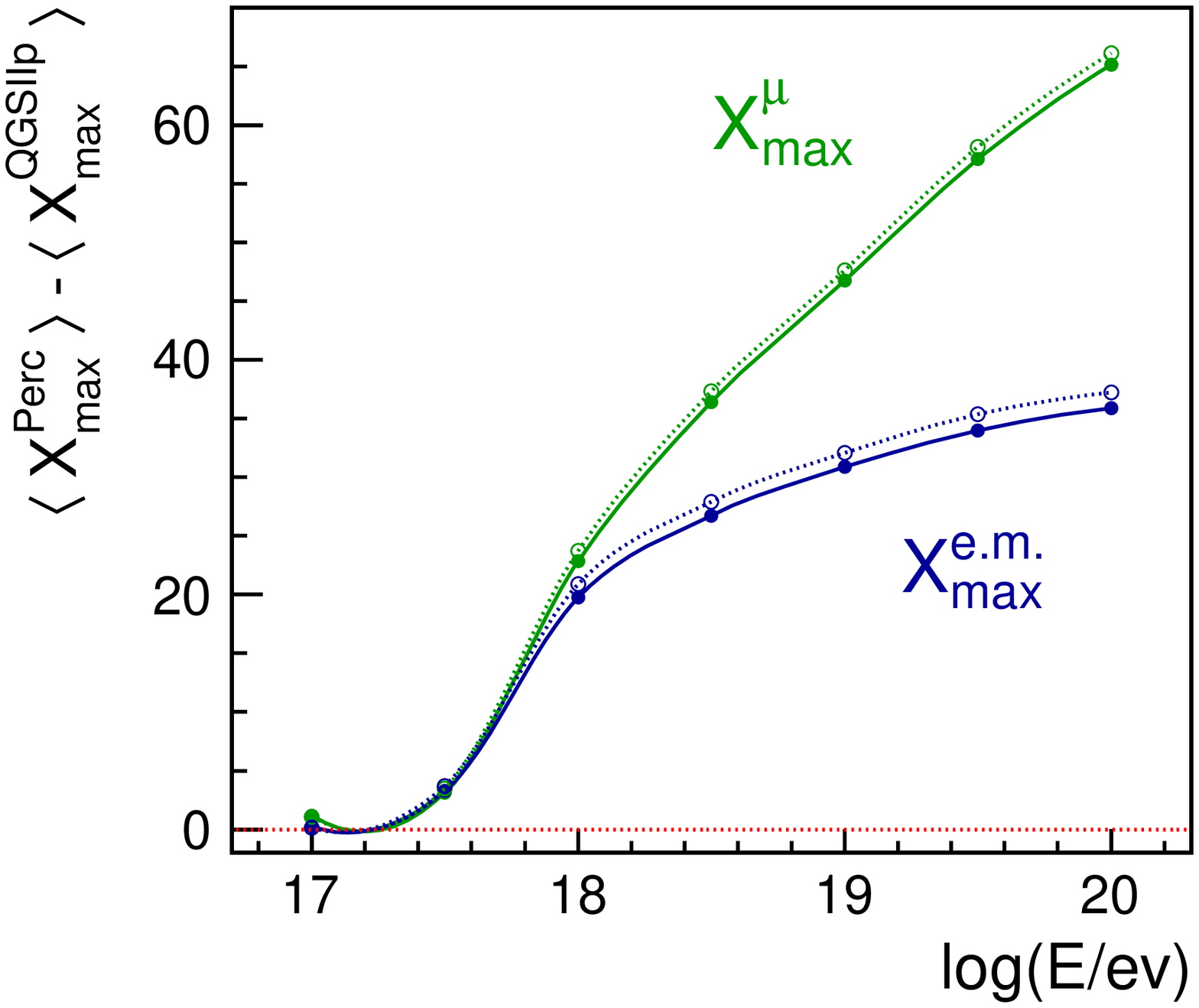}
\includegraphics[width=0.45\textwidth]{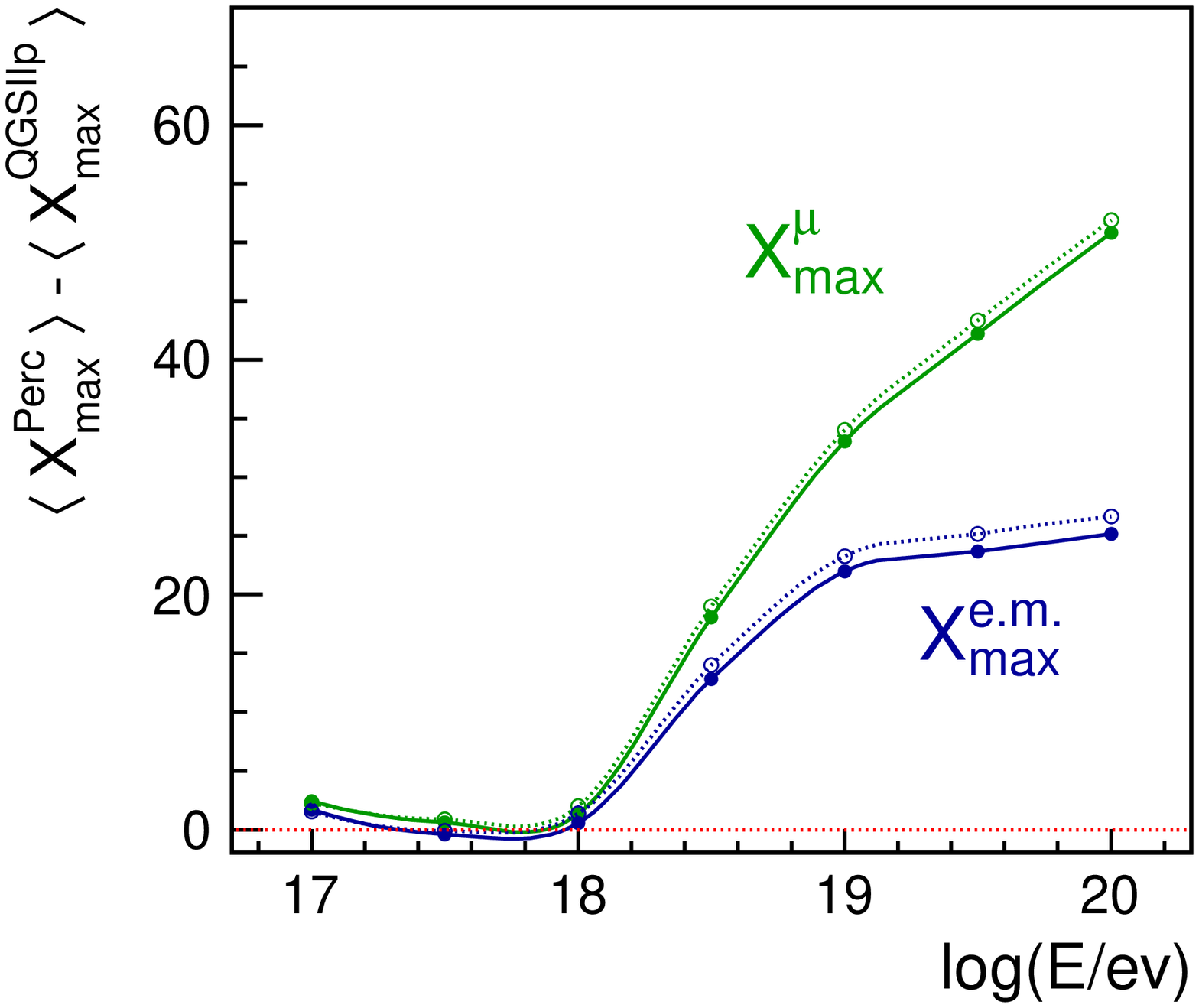}
\caption{Difference between the average depth of maximum of the electromagnetic and muonic longitudinal 
profiles $X^{\rm e.m.}_{\rm max}$ and $X^\mu_{\rm max}$ respectively, as obtained in proton-induced shower
simulations with ($\left< X^{\rm perc}_{\rm max}\right>$) and without ($\left< X^{\rm perc}_{\rm max}\right>$) 
percolation effects implemented. The effect of the percolation energy scale 
(see text for explanation) is shown with dashed lines. Left panel: an uniform density profile for the proton 
is assumed. Right panel: a Gaussian profile is assumed.}
\label{fig:xmax} 
\end{center}
\end{figure}

The effect of percolation on the depth of maximum of the electromagnetic and Muon Production Depth profile \cite{USPVmu,TranspModel}, $X^{\rm e.m.}_{\rm max}$ and $X^{\mu}_{\rm max}$ respectively, was also studied. 
The results are displayed in Fig.~\ref{fig:xmax} as the difference between the average depth of maximum in showers 
with percolation and the corresponding average depth
in standard showers, assuming an uniform density profile for the proton (left panel) and for a Gaussian profile 
(right panel). In both cases the difference in the depth of maximum can be seen to increase with primary energy 
when percolation is effective for both $X^{\rm e.m.}_{\rm max}$ and $X^{\mu}_{\rm max}$. This is easily understood 
as the string percolation model predicts collisions with lower multiplicity and faster particles on average, which transport 
more energy in the forward region of the collison, pushing both $X^{\rm e.m.}_{\rm max}$ and $X^\mu_{\rm max}$ deeper into the atmosphere
than in the case of standard showers without percolation effects. 
At $E = 10^{19}$ eV the value of $X^{\rm e.m.}_{\rm max}$, typically observed in UHECR experiments with the fluorescence detector, 
is $\sim 22\--31\ {\rm g\ cm^{-2}}$ deeper in proton-induced showers with percolation effects than in standard proton-induced showers 
simulated with QGSJ{\textsc{et}}-II. Although the effect is not very large, it may alter the interpretation of UHECR data 
in terms of composition of the primary particles \cite{AugerFD}.   
The depth of the muonic longitudinal profile at production 
$X^{\mu}_{max}$ is also increased by $\sim 33\--47\ {\rm g\ cm^{-2}}$ with respect to that obtained in standard showers. 
Interestingly, the difference between $X^{\mu}_{max}$ in showers with and without percolation increases continuously with the primary energy 
while the corresponding difference for $X^{\rm e.m.}_{\rm max}$ saturates at the highest energies. 
This indicates that the hardening of the energy spectra and the increase of strange particles and baryons, 
have a different impact on the electromagnetic and the hadronic shower components. Moreover, notice that the probability of producing 
a given type of particle saturates quickly after the percolation transition (see Fig.~\ref{fig:pquarkperc}), while the increase of the string rapidity 
due to string fusion, continues to grow steadily (see Fig.~\ref{fig:stringYincrease}).


The impact of a lower energy scale as predicted in the percolation model is also shown in Fig.~\ref{fig:xmax}, although the effect is 
very small. $X^{e.m.}_{\rm max}$ increases by only $\sim 2\ {\rm g\ cm^{-2}}$. 
The small effect of the energy scale on the $X_{max}$, contrary to what happened with the number of muons, is a consequence of the 
logarithmic $X_{max}$ evolution with the primary energy.

The results on the electromagnetic $X^{e.m.}_{\rm max}$ and the muonic $X^{\mu}_{max}$ provide a very distinct experimental signature which could in principle be searched in hybrid experiments such as Pierre Auger Observatory or the Telescope Array Experiment. These experiments combine two independent detection techiques to investigate EAS: Fluorescence Detectors and Surface Detectors. The Fluorescence Detectors \cite{HiResFD,AugerFD,TAFD}, observe the longitudinal shower development, in particular $X^{e.m.}_{max}$, by detecting the light emmited by the nitrogen molecules as the shower secondary particles (mainly electrons) travel through the atmosphere. The surface detector sample the shower secondary charged particles that arrive at ground. Amongst these particles are the muons, which can be used, along with their timing information, to reconstruct the Muon Production Depth \cite{MPDAuger}, giving access to $X^{\mu}_{max}$. 

Note that in the percolation scenario the evolution of both maxima with the primary energy is different. Therefore, the consistency between the interpretation of the measurements of $X^{e.m.}_{max}$ and $X^{\mu}_{max}$ in terms of mass composition is not possible for a hadronic interaction model without percolation.

\section{Conclusions}
\label{sec:Conclusions}

In this paper we have shown that the string percolation phenomenon is expected 
to appear at energies around $\sqrt{s}\sim 30-40$ TeV, i.e, between the LHC energies 
and those probed in UHECR-induced showers collected at the Pierre Auger Observatory. 
The transition energy depends on the assumptions for the proton density profile.

We have shown that string percolation induces an enhancement in the production of strange mesons and the total number of baryons, 
providing a mechanism to increase the muon content in extensive air showers. Moreover, string percolation would  
harden the particle energy spectrum in hadronic interactions and consequently, due to energy conservation, 
it would reduce the multiplicity of the highest energy hadronic interactions.

The effect of string percolation on shower observables was investigated. Percolation leads to a reduction of the energy flow 
into the electromagnetic component of the air shower, thus causing an energy shift if the shower should be interpreted with an energy 
scale with no percolation. 
The natural increment of the energy on the hadronic cascade, caused by percolation, should increase the muon content by up to $\sim 50\%$ 
relative to standard proton-induced extensive air showers at $E = 10^{19}$ eV.

The depth of the maximum of the electromagnetic shower longitudinal profile displays a transition around the percolation energies towards deeper values. The reduction of multiplicity and the hardening of the particle spectrum, effectively slows down the growth of the hadronic shower causing the depth of muonic component maximum to also become deeper.

One of the predictions of the percolation scenario is that the electromagnetic and muon production shower maxima should have different evolutions with the primary energy for a fixed mass composition. This could be experimentally checked by combining the information of the Fluorescence Detectors, which measure the electromagnetic $X^{e.m.}_{max}$, with the one provided by the Surface Detectors, which can be sensitive to the muon production maximum, $X^{\mu}_{max}$.

\section*{Acknowledgments}
\label{sec:Acknowledgments}

We would like to thank to C. Dobrigkeit, S. Andringa, L. Apolin\'ario and F. Diogo for careful reading the manuscript.
This work is partially funded by Funda\c{c}\~ao para a Ci\^encia e Tecnologia (CERN/FP123611/2011 and SFRH/BPD/73270/2010),
and fundings of MCTES through POPH-QREN-Tipologia 4.2, Portugal, and European Social Fund.
We also thank Xunta de Galicia (INCITE09 206 336 PR) and
Conseller\'\i a de Educaci\'on (Grupos de Referencia Competitivos --
Consolider Xunta de Galicia 2006/51); Ministerio de Ciencia e Innovaci\'on
(FPA2007-65114, FPA 2008-01177, FPA2010-18410, FPA2011-22776,
FPA2012-39489 and Consolider CPAN - Ingenio 2010) and Feder Funds, Spain.

\bibliographystyle{elsarticle-num}
\bibliography{Bib-strangeness}

\end{document}